\documentclass[nonacm,sigplan,10pt]{acmart}
\acmYear{2023}
\acmISBN{} %
\acmDOI{} %
\startPage{1}

\setcopyright{none}

\usepackage{booktabs}   %
\usepackage{subcaption} %

\usepackage{hyperref}
\usepackage{booktabs} %
\usepackage{algorithmicx}
\usepackage[ruled]{algorithm}
\usepackage{algpseudocode}
\usepackage{bold-extra}
\usepackage{setspace}
\usepackage{syntax}
\usepackage{balance}
\usepackage{xspace}
\usepackage{proof}
\usepackage{amsmath}
\usepackage{tikz}
\usetikzlibrary{positioning,matrix,arrows,arrows.meta,backgrounds}
\usepackage{stmaryrd}
\usepackage[T1]{fontenc}
\usepackage{multicol}
\usepackage[inline]{enumitem}
\usepackage{xcolor}
\usepackage{ifthen}
\usepackage{siunitx}
\usepackage{microtype}

\definecolor{stringColor}{rgb}{0.8,0.1,0.1}
\usepackage{listings}
\lstdefinelanguage{LF}{
  keywords={msec, sec, timer, startup, shutdown, state, main, actor, handler, reaction, preamble, target, reactor, composite, trigger, input, output, constructor, new, action, clock, logical, physical, after, import, from, private, public, method, time, interleaved, extends},
  keywordstyle=\color{black}\bfseries,
  ndkeywords={class, export, boolean, throw, implements, this, int, if, else, float},
  ndkeywordstyle=\color{darkgray}\bfseries,
  identifierstyle=\color{black},
  sensitive=false,
  comment=[l]{//},
  morecomment=[s]{/*}{*/},
  commentstyle=\color{black}\ttfamily,
  stringstyle=\color{black}\ttfamily,
  morestring=[b]',
  morestring=[b]"
}
\newcounter{myctr}

\input{support/macros}

\newcommand{\lf}[0]{{\sc Lingua Franca}\xspace}
\newcommand{\lfshort}[0]{LF\xspace}

\newcommand{\lfheader}[0]{\textsc{Lingua Franca}\xspace}
\newcommand{\lfdef}[0]{\lf (\lfshort)}

\makeatletter
\def\@ACM@checkaffil{%
  \if@ACM@countrypresent\else
    \ClassWarningNoLine{\@classname}{No country present for an affiliation}%
  \fi
}
\makeatother

\begin{document}
\title[High-Performance Deterministic Concurrency using \lfheader]{High-Performance Deterministic Concurrency using \lfheader}

\author{Christian Menard}
\orcid{0000-0002-7134-8384}
\email{christian.menard@tu-dresden.de}
\affiliation{%
  \institution{TU Dresden, Germany}
}

\author{Marten Lohstroh}
\orcid{0000-0001-8833-4117}
\email{marten@berkeley.edu}
\affiliation{%
  \institution{UC Berkeley, USA}
}

\author{Soroush Bateni}
\orcid{0000-0002-5448-3664}
\email{soroush@berkely.edu}
\affiliation{%
  \institution{UC Berkely, USA}
}

\author{Matthew Chorlian}
\orcid{0000-0003-4102-7181}
\email{mattchorlian@berkeley.edu}
\affiliation{%
  \institution{UC Berkeley, USA}
}

\author{Arthur Deng}
\email{langxing.deng@berkeley.edu}
\affiliation{%
  \institution{UC Berkeley, USA}
}

\author{Peter Donovan}
\orcid{0000-0003-3374-0753}
\email{peterdonovan@berkeley.edu}
\affiliation{%
  \institution{UC Berkeley, USA}
}

\author{Clément Fournier}
\orcid{0000-0002-5661-3004}
\email{clement.fournier@tu-dresden.de}
\affiliation{%
  \institution{TU Dresden, Germany}
}

\author{Shaokai Lin}
\orcid{0000-0001-6885-5572}
\email{shaokai@berkeley.edu}
\affiliation{%
  \institution{UC Berkeley, USA}
}

\author{Felix Suchert}
\orcid{0000-0001-7011-9945}
\email{felix.suchert@tu-dresden.de}
\affiliation{%
  \institution{TU Dresden, Germany}
}

\author{Tassilo Tanneberger}
\orcid{0000-0002-3196-7869}
\email{tassilo.tanneberger@tu-dresden.de}
\affiliation{%
  \institution{TU Dresden, Germany}
}

\author{Hokeun Kim}
\orcid{0000-0003-1450-5248}
\email{hokeun@hanyang.ac.kr}
\affiliation{%
  \institution{Hanyang University, South Korea}
}

\author{Jeronimo Castrillon}
\orcid{0000-0002-5007-445X}
\email{jeronimo.castrillon@tu-dresden.de}
\affiliation{%
  \institution{TU Dresden, Germany}
}

\author{Edward A. Lee}
\orcid{0000-0002-5663-0584}
\email{eal@berkeley.edu}
\affiliation{%
  \institution{UC Berkeley, USA}
}

\renewcommand{\shortauthors}{Menard \& Lohstroh et al.}

\begin{abstract}
Actor frameworks and similar reactive programming techniques are widely used
for building concurrent systems. They promise to be efficient and scale
well to a large number of cores or nodes in a distributed system. However,
they also expose programmers to nondeterminism, which often makes implementations
hard to understand, debug, and test. The recently proposed
reactor model is a promising alternative that enables efficient deterministic
concurrency. In this paper, we show that determinacy does neither imply a
loss in expressivity nor in performance. To show this, we evaluate \lfdef, a reactor-oriented
coordination language that equips mainstream programming languages
with a concurrency model that automatically takes advantage of opportunities
to exploit parallelism that do not introduce nondeterminism.
Our implementation of the Savina benchmark suite demonstrates that, in terms of
execution time, the runtime performance of \lfshort programs even exceeds
popular and highly optimized actor frameworks. We compare against Akka and CAF,
which \lfshort outperforms by $1.86x$ and $1.42x$, respectively.

\end{abstract}



\maketitle

\section{Introduction}\label{sec:intro}

Theoreticians working on programming language semantics have long understood the value of determinism as well as the expressive power of nondeterminism in programming languages. In practice, however, today, nondeterminism creeps into programming languages and frameworks not to benefit from its expressiveness, but rather because of a widespread perception that it is needed to get good performance on parallel hardware. We show in this paper that it is not necessary to sacrifice determinism to achieve performance. We do this by focusing on actor frameworks, which have proved popular and successful in many very demanding applications, but admit nondeterminism that is often not actually needed by their applications.

Exploiting parallel hardware such as multicore machines to improve performance is only possible when programs expose concurrency.
Common abstractions for concurrency
include threads~\cite{cooper1988c},
remote procedure calls~\cite{nelson1981remote},
publish-subscribe~\cite{Eugster:03:PubSub},
service-oriented architectures~\cite{PerreyLycett:03:SOA}, and
actors~\cite{Agha:97:ActorComputation,hewitt2010actor}.
Each of these models has its own merits, but they all
introduce nondeterminism: situations where, for a given state and input, the behavior of a program is not uniquely defined.
While nondeterminism can be useful in some applications, most programming tasks benefit from more repeatable behavior.
Deterministic programs are easier to understand, debug, and test (for each test vector, there is one known-good response).
For nondeterministic programs, problematic behaviors might be harder to discover because they may only occupy a small
fraction of the state space~\cite{KranzlmullerDieter:02:Nondeterminism}.
And reproducing failures can be extremely
hard~\cite{Lee:06:Threads,Liu:21:Threads:PLDI,musuvathi:08:finding} because they
might occur only when the system is under a specific amount of load~\cite{Sen:RaceFuzzer}.

Determinism is a subtle concept~\cite{Lee:21:Determinism}.
Here, we focus on a particular form of determinism for programs, where a program is deterministic if, given the same inputs, it always produces the same outputs.
This definition does not require that operations be performed in a particular order, and therefore is not at odds with concurrency and parallel execution.
It is possible, but often not easy, to achieve this form of determinism even when using nondeterministic abstractions such as threads, actors, and asynchronous remote procedure calls.
For simple enough programs, such as a chain of actors, if communication is reliable, then execution will be deterministic.
Some of the benchmarks we compare against in this paper are deterministic in this way.
As we will show, however, even slightly more complex communication structures result in nondeterminism that can be difficult to correct.

In this paper, we evaluate a language-based coordination approach to specifying
concurrent software that preserves determinism by default and only admits
nondeterminism when explicitly introduced by the programmer. 
The coordination language
\lfdef~\cite{LohstrohEtAl:21:Towards}, which is based on a concurrent model of computation called 
reactors~\cite{Lohstroh:2019:CyPhy, Lohstroh:EECS-2020-235}, achieves this
by analyzing program structure and ensuring that data dependencies are observed correctly at runtime. 
An \lfshort program defines reactive software components called ``reactors''
and provides operators to compose them hierarchically (through containment) and bilaterally (via connections).
Because the language supports both deterministic and nondeterministic concurrency, it provides a fertile ground for exploring the impact of determinism on performance.

The semantics of the deterministic subset of \lfshort can be thought of as a deterministic variant of actors~\cite{Agha:97:ActorComputation,hewitt2010actor,LohstrohLee:19:Actors}.
We show in this paper that it delivers performance
comparable to popular nondeterministic realizations of actors on parallel hardware.
Like Akka~\cite{AkkaAction2016} and CAF~\cite{chs-rapc-16}---the frameworks we compare against---\lfshort
orchestrates the execution of chunks of code written in conventional programming languages, allowing programmers to rely on the languages, libraries, and tools that they are comfortable with.
Unlike the frameworks we compare against, \lfshort is polyglot.
It currently supports C, C++, Python, TypeScript, and Rust. This paper focuses
on the runtime performance of the C++ target, which, as a core contribution of
this paper, has been optimized to efficiently exploit concurrency on parallel
hardware. Earlier work~\cite{LohstrohEtAl:21:Towards} has only reported preliminary 
performance indications of \lfshort based on its C target, which is 
predominantly aimed at running on embedded systems.

At the core of \lfshort's concurrency model
is a logical model of time that gives a clear notion of simultaneity and
avoids deadlocks using dependency analysis based on causality 
interfaces~\cite{Lee:05:CausalityInterfaces}. 
It is this timed semantics that enables efficient deterministic concurrency in \lfshort.
However, the benchmarks we compare against were created to evaluate actor frameworks,
which have no temporal semantics. None of the benchmarks take advantage of the time-related features of \lfshort; the temporal semantics is only used to deliver determinism.

\paragraph{Contributions.}
We show that the reactor-oriented paradigm as implemented in \lf enables efficient exploitation of parallel hardware
without relinquishing determinism.
For this, we explain the mechanisms through which
\lfshort programs expose concurrency;
we present a language extension that allows for the definition of scalable programs;
and we introduce an optimized C++ runtime for \lfshort that enables efficient parallel execution.
We further present an extensive evaluation based on the Savina benchmark suite~\cite{savina}, showing
that our \lfshort runtime outperforms
Akka and CAF by \(1.86x\) and \(1.42x\), respectively.

\paragraph{Outline.}

We first motivate our work (Section~\ref{sec:motivation}) and then introduce \lfshort
(Section~\ref{sec:intro-lf}). We then go into detail about the concurrency in
\lfshort and introduce our optimized C++ runtime (Section~\ref{sec:concurrency}). 
Next, we report benchmark results
(Section~\ref{sec:evaluation}), discuss related work (Section~\ref{sec:related}), and
conclude (Section~\ref{sec:conclusion}).

\section{Motivation}\label{sec:motivation}

The actor model is widely accepted and deployed in production for its promise to
allow programmers to easily express concurrency, provide high execution
performance, and scale well to large datasets and complex applications.
Moreover, in contrast to thread-based programs, actor semantics prevents \emph{low-level} data races.
However, like most message passing paradigms, actors expose the programmer to
nondeterminism in the form of \emph{high-level} data
races~\cite{TorresLopez2018}, a problem that becomes considerably challenging to
manage as the complexity of a program grows.

\begin{figure}[tb]
  \tikzstyle{actor}=[circle, draw, minimum size=4em, inner sep=0]
  \centering
  \begin{minipage}{0.4\linewidth}
  \subcaptionbox{Deposit and Withdrawal sent by different users.\label{fig:actor-example-2users}}{
    \begin{tikzpicture}[->,>=Latex,thick, node distance=6em]
      \sffamily
      \scriptsize
      \node[] (dummy) {};
      \node[actor, above=0.8em of dummy] (u1) {User A};
      \node[actor, below=0.8em of dummy] (u2) {User B};
      \node[actor, right=of dummy] (a) {Account};
      \draw [->] (u1) to [out=0,in=135] node[above,inner sep=5pt]{Deposit} (a);
      \draw [->] (u2) to [out=0,in=-135] node[below,inner sep=5pt]{Withdrawal} (a);
    \end{tikzpicture}
  }
  \end{minipage}\hfill
  \begin{minipage}{0.59\linewidth}
    \centering
  \subcaptionbox{Deposit and Withdrawal sent by same user.\label{fig:actor-example-1user}}{
    \begin{tikzpicture}[->,>=Latex,thick, node distance=5em]]
      \sffamily
      \scriptsize
      \node[actor] (u) {User};
      \node[actor, right=of u] (a) {Account};
      \draw [->] (u) to [out=20,in=160] node[above]{Deposit} (a);
     \draw [->] (u) to [out=-20,in=-160] node[below]{Withdrawal} (a);
    \end{tikzpicture}
  }\hfill
  \subcaptionbox{Withdrawal sent via a proxy.\label{fig:actor-example-proxy}}{
    \begin{tikzpicture}[->,>=Latex,thick, node distance=5em]]
      \sffamily
      \scriptsize
      \node[actor] (u) {User};
      \node[right=of u] (dummy) {};
      \node[actor, below=1ex of dummy] (p) {Proxy};
      \node[actor, right=of dummy] (a) {Account};
      \draw [->] (u) to [out=10,in=170] node[above]{Deposit} (a);
      \draw [->] (u) to [out=-30,in=170] node[below, inner sep=7pt, near start,xshift=-4pt]{Withdrawal} (p);
      \draw [->] (p) to [out=10,in=-150] node[below, inner sep=5pt,xshift=6pt]{Withdrawal} (a);
      \end{tikzpicture}
  }
  \end{minipage}
  \vspace{-8pt}
  \caption{Example actor programs that may expose nondeterministic behavior.}
  \label{fig:actor-example}
\end{figure}
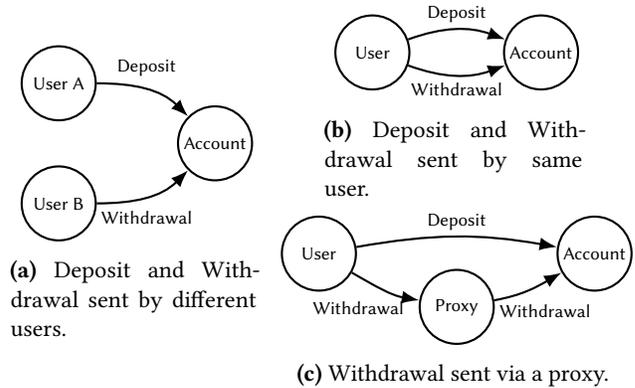

\begin{figure*}[tb]
  \subcaptionbox{Simple \lfshort implementation.\label{fig:lf-example-2users}}{
    \includegraphics[scale=0.57]{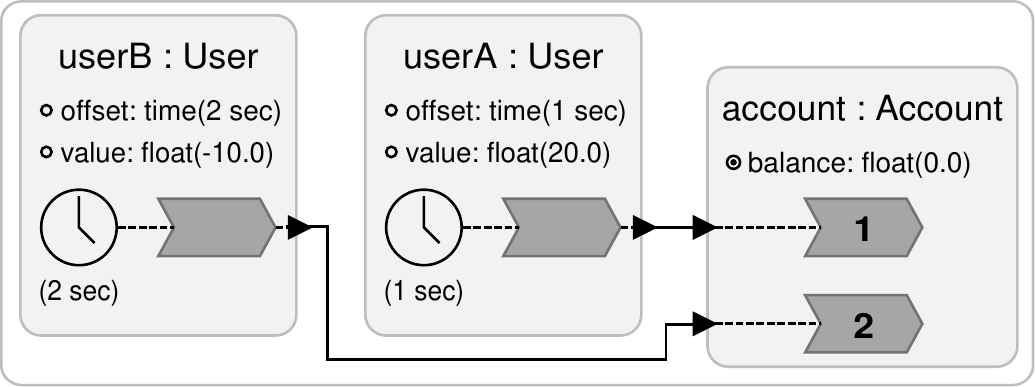}
  }
  \subcaptionbox{Adding a proxy reactor.\label{fig:lf-example-proxy}}{
    \includegraphics[scale=0.57]{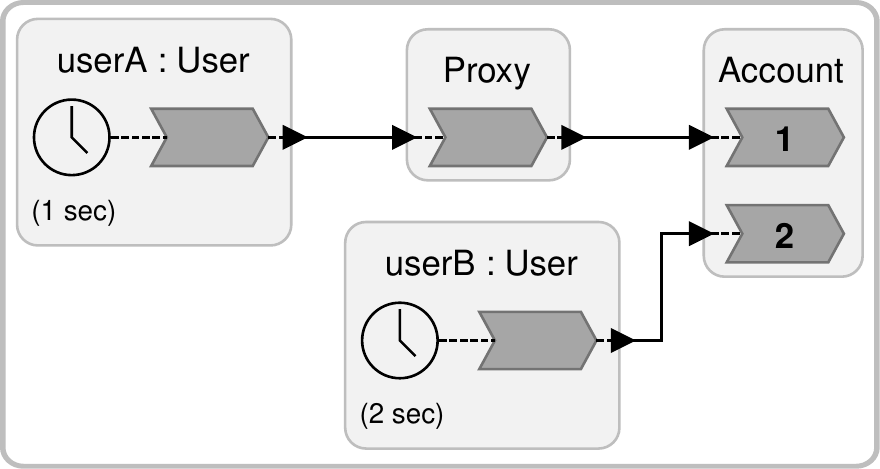}
  }
  \subcaptionbox{Adding a proxy introducing a logical delay.\label{fig:lf-example-proxy-delay}}{
    \includegraphics[scale=0.57]{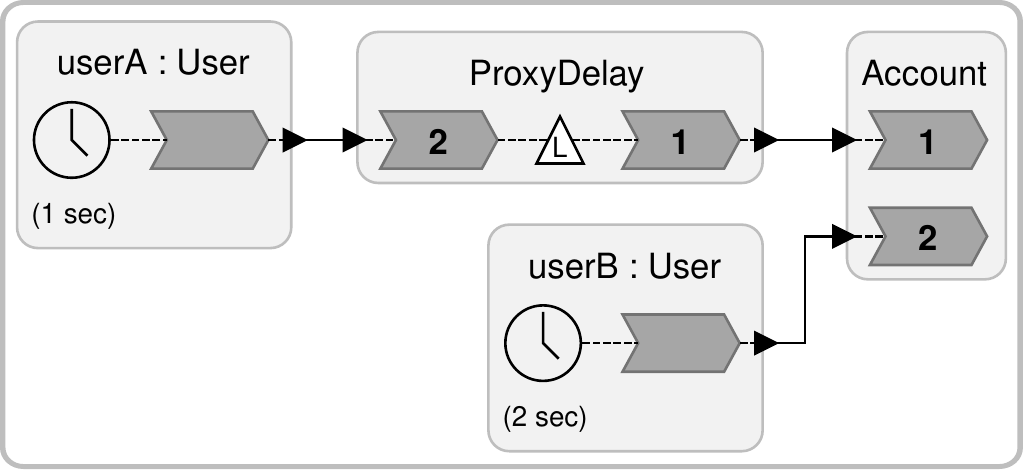}
  }
  \vspace{-6pt}
  \caption{Example \lfshort implementation of the actor program shown in Figure~\ref{fig:actor-example-2users} and variations thereof.}
  \label{fig:lf-examples}
  \vspace{-6pt}
\end{figure*}

Consider the simple example in Figure~\ref{fig:actor-example-2users}. The \texttt{Account}
actor manages the balance of a bank account that two users interact with.
\texttt{User A} sends a deposit message increasing the account's balance and
\texttt{User B} sends a
withdrawal message decreasing the account's balance. If we assume that the balance
is initialized to 0 and the account only grants a withdrawal if the resulting
balance is not negative, then there are two possible behaviors. If A's message
is processed first, then the withdrawal is granted to B. If B's message is
processed first, then the withdrawal is denied. The actor model assigns no
meaning to the ordering of messages.
Therefore, there is no well-defined correct behavior for this example.

The reader may object that for an application like that of Figure~\ref{fig:actor-example-2users},
the order of transactions is intrinsically nondeterministic, and any additional nondeterminism
introduced by the software framework is inconsequential.
However, if we focus on testability, we see that even identical inputs can yield different
results, making testing more difficult.
If we focus on consistency, the problem that different observers of the same events may
see different behaviors becomes problematic.
In databases, it is common to assign time stamps to external inputs
and to then treat those timestamps as a semantic property of the inputs and define the behavior
of the database relative to those time stamps.
We adopt this perspective in this paper, and rely on the definition of determinism given by \citet{Lee:21:Determinism}:
``determinism is a property of models, not of physical realizations,''
and
``A model is deterministic if given all the \emph{inputs} that are provided to the model, the
model defines exactly one possible \emph{behavior}.''
If we define ``inputs'' in Figure~\ref{fig:actor-example-2users} to be time-stamped user queries
and ``behavior'' to be the sequence of actions taken by the Account, then it is reasonable to demand determinism.

Consider Figure~\ref{fig:actor-example-1user}, which has only one user.
Even if this one user first sends a deposit and then a withdrawal message, 
the actor model does
not guarantee that the receiving actor sees and processes the incoming messages
in this order. While some actor frameworks, e.g., Akka and Erlang, guarantee in-order message
delivery, others, e.g., AmbientTalk~\cite{vanCutsem2014ambienttalk},
expressly do not.
Yet, even if the framework guarantees point-to-point in-order message delivery,
this property is not transitive. If we add a \texttt{Proxy}, as shown in
Figure~\ref{fig:actor-example-proxy}, then we cannot make any assumptions about the
order in which \texttt{Account} receives messages. This example further illustrates that
composing actors can have unexpected side effects.

Consequently, implementing solutions to practical concurrency problems with actors can be
challenging. Even seemingly simple concurrency problems like the one discussed
above require high programming discipline, and solutions are typically difficult
to maintain and tend to lack modularity. In addition, the inherent nondeterminism of actor frameworks makes
it hard to verify such solutions. Erroneous behavior might only occur
in a fraction of executions, and thus integration tests cannot reliably detect
such ``Heisenbugs''~\cite{musuvathi:08:finding}.

In a recent study, \citet{Bagherzadeh2020} analyzed bugs in Akka programs that were
discussed on StackOverflow or GitHub and determined that 14.6\% of the bugs are
caused by races. This makes high-level races the second most common
cause of bugs in Akka programs after errors in the program logic. In a similar
study of 12 actor-based production systems, \citet{Hedden2018} determined
that 3.2\% of the reported bugs were caused by bad message ordering, 4.8\% of
bugs were caused by incorrect coordination mechanisms, 4.8\% were caused by
erroneous coordination at shutdown, and 2.4\% of bugs were caused by erroneous
coordination at startup.
Note that these numbers only cover \emph{known} bugs in
their studied projects and, as noted by the authors, the majority of the
reported message ordering bugs belonged to the Gatling project because it
already incorporated a debugging tool called Bita~\cite{Tasharofi2013} that is
designed to detect such bugs. We suspect that there are more undetected bugs in
projects that do not use specialized debugging tools.

The actor community has addressed the inherent nondeterminism of actors and the
resulting bugs by introducing better tools for analyzing and debugging actor
programs. This includes TransPDOR~\cite{Tasharofi2012},
Bita~\cite{Tasharofi2013}, Actoverse~\cite{Shibanai2017},
iDeA~\cite{Mathur2018}, CauDEr~\cite{Lanese2018}, and Multiverse
debugging~\cite{TorresLopez2019}. While these are valuable solutions, we argue that a
programming model for expressing concurrent programs should provide
deterministic semantics by default and allow the programmer to %
introduce nondeterminism only where it is desired and understood to do no harm.
In such cases, the aforementioned tools for
nondeterministic behavior can still be utilized to debug the implementation.

There are a number of ways to achieve deterministic concurrency,
including Kahn process networks~\cite{Kahn:74:PN,Kahn:77:PN},
many flavors of dataflow models~\cite{Dennis:74:Dataflow,NajjarEtAl:99:Dataflow,LeeMatsikoudis:09:Dataflow},
physically asynchronous, logically synchronous models~\cite{ShaEtAl:09:PALS},
synchronous-reactive languages~\cite{Benveniste:91:Synchronous,EdwardsLee:03:Synchronous},
and discrete-event systems~\cite{Zeigler:76:DEVS,Cassandras:93:DE,LeeZheng:07:SRDECT,DBLP:conf/fdl/EdwardsHL20}.
\citet{LohstrohEtAl:21:Towards} compare the reactor model to each of these,
showing that it has many of their best features and fewer of their pitfalls.
\lf builds on this reactor model because it is more expressive than some of the alternatives (e.g., Kahn networks)
and is stylistically close to actors, which have proven effective in practice.
In this paper, we show that the resulting determinism does not incur a performance penalty,
but on the contrary, helps to achieve improved performance in most cases.

\section{Introduction to \textbf{\textsc{Lingua Franca}}}\label{sec:intro-lf}

\lfdef{} builds on the relatively new reactor-oriented programming paradigm.
Intuitively, we can describe reactors as deterministic actors with a
discrete event execution semantics and explicitly declared ports and
connections. A logical timeline is used to order events and ensure a
deterministic execution.
As a polyglot language, \lfshort incorporates
code in a target programming language to implement the logic of each component.
\lfshort itself is only concerned with the coordination aspect of a program.

In this section, we introduce the core concepts of reactors and \lfshort. Note
however, that a full discussion of \lfshort including its syntax and tooling is
beyond the scope of this paper. Instead, we base our discussion on \lfshort's
diagrammatic representation of programs%
which gets synthesized from \lfshort source code automatically~\cite{Isola22}.
A complete introduction to \lfshort's textual syntax is given by \citet{LohstrohEtAl:21:Towards}.

\subsection{\lfshort by Example}

Figure~\ref{fig:lf-example-2users} shows an \lfshort implementation of the
deposit/withdrawal example in Figure~\ref{fig:actor-example-2users}.
The program
is assembled from three \textbf{reactor instances} \texttt{userA}, \texttt{userB} and
\texttt{account}, shown as light gray boxes in the diagram. Note that
\texttt{userA} and \texttt{userB} are instances of the same \textbf{reactor class}
\texttt{User} and hence share the same structure and functionality.
In the diagram, black triangles denote \textbf{ports}. In this example, both users have
an output port which is connected to a respective input port at the account.
These ports and connections allow the users to send requests to the account.

In \lfshort, all computation is performed in reactive code segments called \textbf{reactions} that are implemented in an arbitrary target language. In the diagram,
reactions are represented by dark gray chevrons. All reactions
must explicitly declare their triggers, other dependencies and potential
effects.%
In the example in Figure~\ref{fig:lf-example-2users}, both users define a
reaction that is triggered by a \textbf{timer}. Timers are an \lfshort construct used 
to produce events in regular intervals or once at a specific time.
The timer of \texttt{userA} is configured to trigger an event one
second after program startup; the timer of \texttt{userB} is configured to
trigger an event two seconds after program startup. The corresponding reactions
simply send a deposit or withdrawal request by setting the user's output port.

The \texttt{Account} reactor defines two reactions, one for each of its inputs. Both
reactions will simply try to apply the requested change to the balance, which is stored in a state variable local to the reactor instance. Note
that reactors may define arbitrary state variables which are accessible by all
their own reactions (which does not include reactions of contained reaction). 
In addition to state variables, reactors may also define parameters which can be
set at instantiation. This mechanism allows the \texttt{User} reactor to be
reusable, as the precise time at which the timer triggers (\texttt{offset}) and
the amount to withdraw or deposit (\texttt{value}) are configurable at instantiation time.

The reader might notice that the separated reactions in \texttt{account}
duplicate logic and are not a practical solution, in particular if there are many
users. We choose this representation to keep our
exposition simple and avoid a detailed discussion of \lfshort syntax. Indeed, in
\lfshort a single reaction can bind to an arbitrary number of upstream ports.

When executed, the program will wait for 1 second before triggering the timer
of the \texttt{userA} reactor and invoking its reaction. The event produced by this reaction will
trigger reaction 1 in \texttt{account}, which is invoked
immediately after the first reaction completes. Two seconds after program startup,
\texttt{userB} will react and subsequently trigger reaction 2 in \texttt{account}.
In this example, the deposit event (+20.0) occurs earlier than the withdrawal
event (-10.0), and hence our execution semantics ensures that the account processes the
deposit event before the withdrawal event, meaning the balance will not become negative.
In a more realistic implementation, the two
users would generate events sporadically and have their reactions 
triggered not by a timer but a physical action (see
Section~\ref{sec:physical-actions}). However, using a timer greatly simplifies our
exposition as we only have to consider a single \emph{logical timeline} along
which events are ordered.
Moreover, such timers can be used to create regression tests that validate program execution
with specific input timings.

Note that even when the two events occur logically simultaneously, meaning that both reactions in the 
\texttt{account} reactor are triggered at the same logical time,
the resulting program will be deterministic.
All reactions within the same tag are executed according to a well-defined precedence relation.
In particular, any reactions within the same reactor are mutually exclusive and executed following the
lexical declaration order of the reactions in \lfshort code.
This order is also reflected by the numbers displayed on the reactions
in the diagrams in Figure~\ref{fig:lf-examples}.
More details on the precedence relation of reactions are given in Section~\ref{sec:parallelism}.

To deliberately change the order in which events occur, a logical delay can be
introduced in the program using a \textbf{logical action}, as shown in
Figure~\ref{fig:lf-example-proxy-delay}. In the diagram, actions are denoted by small white triangles.
In contrast to ports, which allow relaying events logically instantaneously from one reactor to another, logical actions provide a mechanism
for scheduling new events at a later (logical) time. Upon receiving an input, reaction 2 of the \texttt{ProxyDelay} reactor is triggered, which schedules its logical action with a configurable delay.
This creates a new event which, when processed, triggers reaction 1 of the \texttt{ProxyDelay} reactor, which retrieves the original value and forwards it to its output port.

If we assume that a delay of 2 seconds is used for scheduling the logical
action, then the deposit message from \texttt{userA} will only arrive at the
account 3 seconds after startup. Hence the deposit message will be processed
\emph{after} the withdrawal message from \texttt{userB}, causing B's request to be
denied.

It is important to note that all of the discussed examples are deterministic, regardless of
the physical execution times of reactions, as all events are unambiguously
ordered along a single logical timeline.
The physical timing of the events, on the other hand, will be approximate.
The contribution of this paper is to show that such determinism does not necessarily reduce performance
and is also useful for applications that have no need for explicit timing.

\subsection{Logical and physical time}\label{sec:timelines}

All events have an associated \textbf{tag}, 
which is used to order events on a logical timeline at runtime. 
In time-sensitive applications, tags are not purely used for logical ordering but also relate to
physical time. By default, the runtime only processes the events associated with a certain
tag once the current physical time $T$ is greater than the time value of the
tag $t$ ($T>t$).
We say that logical time ``chases'' physical time.
The relationship between physical and logical time in the reactor model gives
logical delays a useful semantics and also permits the formulation of
deadlines.
This timed semantics is particularly useful for software that operates in
cyber-physical systems.
For a more in depth discussion of \lfshort's timed-semantics, the interested reader may refer
to~\cite{LohstrohEtAl:20:LF}.

If an application has no need for any physical time properties, the concurrence
of physical and logical time can be turned off; in this case, the tags are used
only to preserve determinism, not to control timing. Moreover, \lfshort
programmers are not required to explicitly control timing aspects of their
programs. Delays can simply be omitted, for instance when scheduling an action,
in which case the runtime will use the next available tag. In consequence, also
untimed general purpose programs can benefit from the deterministic concurrency
enabled by \lfshort's timed-semantics.

\subsection{Asynchrony and deliberate nondeterminism}\label{sec:physical-actions}

The reactor model distinguishes logical actions and \emph{physical} ones.
A logical action is always scheduled synchronously with a delay relative
to the current tag. A \textbf{physical action} may be scheduled from asynchronous contexts;
its event is assigned a logical time based on the current reading of
physical time. Physical actions are the key mechanism for handling sporadic
inputs originating from physical processes (such as users initiating withdrawal or deposit requests).

The assignment of tags to physical actions is nondeterministic in the sense that it is not defined by the program.
However, once those tags are assigned, for example, to deposit or withdrawal requests by a user, the processing of the events is
deterministic and occurs in tag order.
Hence, the tags assigned to externally initiated events are considered as part of the \emph{input},
and given this input, the program remains deterministic.
This approach draws a clear perimeter around the deterministic and therefore
testable program logic while allowing it to interact with sporadic external
inputs.

Physical actions can also be used within the program itself, for example, to nondeterministically assign a new tag to
a message received from another reactor.
In this usage, physical actions
provide a means for deliberately introducing actor-like nondeterminism into a program.

\section{Efficient Deterministic Concurrency}\label{sec:concurrency}

\lfshort programs are deterministic by default. This property is inherited from
the reactor model that \lfshort implements. \citet{LohstrohEtAl:21:Towards}
explain why reactors behave deterministically. Their argument can be adapted to
the concrete context of the \lf language, but this is beyond the scope of
this paper.
Reactors are also concurrent, and, as we show in this paper, the exposed concurrency is sufficient
for the runtime system to effectively exploit multi-core hardware to where it matches or exceeds the performance of fundamentally asynchronous and nondeterministic actor frameworks.
In this section, we first show exactly how concurrency is exposed and then
describe in more depth how our C++ runtime is implemented and how it utilizes
parallel hardware.

\subsection{Parallelism}\label{sec:parallelism}

The use of statically declared ports and connections as the interfaces between
reactors, as well as the declarations of reaction dependencies, distinguish
reactors from more dynamic models like actors or other asynchronous message
passing frameworks where communication is purely based on addresses.
While the fixed topology of reactor programs is less flexible and limits
runtime adaptation, it also provides two key advantages.
First, it achieves a separation of concerns between the functionality of
components and their composition.
Second, it makes explicit at the interface level which dependencies exist
between components.
As a consequence, a dependency graph can be derived for any composition of reactors.
The dependency graph is an \textbf{acyclic precedence graph (APG)} that organizes all reactions into a partial order that
captures all scheduling constraints that must be observed to ensure that the
execution of a reactor program yields deterministic results.
Because this graph is valid irrespective of the contents of the code that
executes when reactions are triggered, reactions can be treated as a black box.
It is this property that enables the polyglot nature of \lfshort
and exposes the concurrency in the application.

\begin{figure}[tb]
  \centering
  \includegraphics[scale=0.4]{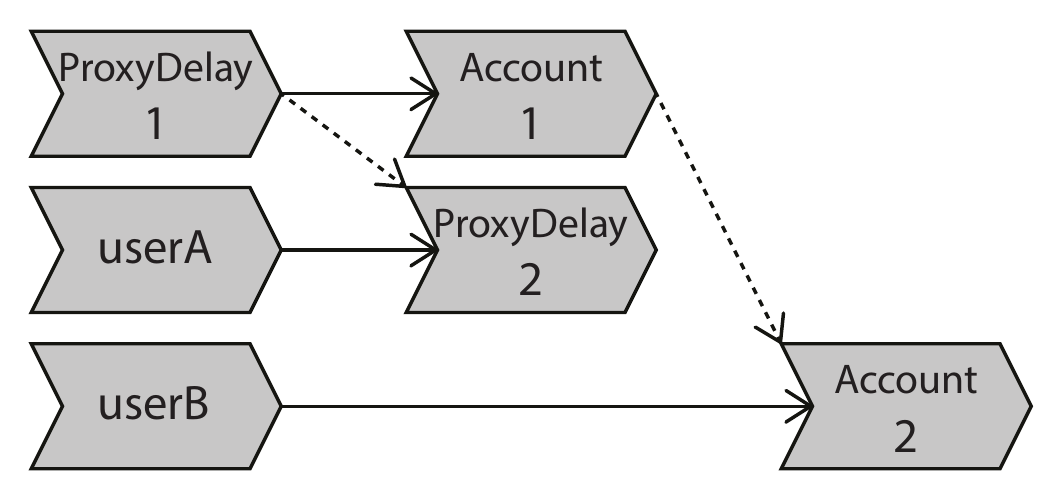}
  \vspace{-8pt}
  \caption{Reaction graph for the program in Figure~\ref{fig:lf-example-proxy-delay}.\label{fig:dependency}}
  \vspace{-10pt}
\end{figure}

Figure~\ref{fig:dependency} shows the dependency graph for the program given in Figure~\ref{fig:lf-example-proxy-delay}.
The solid arrows represent dependencies that arise because one reaction (possibly) sends data to the other.
The dashed arrows represent dependencies that arise because the two reactions belong to the same reactor.
Analogous to the behaviors of actors,
reactions of the same reactor are mutually exclusive.
The execution order is well-defined and given by the lexical declaration order of the reactions in \lfshort code.
This order is also indicated by the numbers in the reaction labels in Figure~\ref{fig:lf-examples}.

The dependency graph precisely defines in which order reactions need to be executed.
Independent reactions may be executed in parallel without breaking determinism.
For instance, the APG in Figure~\ref{fig:dependency} tells us that reaction 1 of
\texttt{ProxyDelay} and the reactions of \texttt{userA} and \texttt{userB} can
all execute in parallel.
Note that the dependency graph is required to be acyclic as any cycle would violate causality.
The \lfshort compiler ensures that a valid program has an acyclic dependency graph.
Any dependency cycles in \lfshort programs can be resolved by introducing a logical action and using it to schedule a new event at a future tag.

Since in \lfshort all dependencies are statically declared, there is a lack of
runtime agility compared to actors and similar models.
The reactor model compensates this with \textbf{mutations} that support runtime
adaptations of the reactor topology and the implied dependency graph.
However, \lfshort does not fully implement mutations yet and a discussion of
mutations is beyond the scope of this paper.

\subsection{Scalable Connection Patterns}\label{sec:connections}

Creating individual reactor instances, ports and connections becomes tedious for
larger programs.
To address this problem, we introduce an extension to the \lfshort
syntax that allows to create multiple ports or reactor instances at once.
Further, we introduce an overloading
of \lfshort's connection operator to create multiple connections over
multiports and banks at once.
This mechanism allows realizing various complex connection patterns in a
single line of code and, as it is fully parameterizable, allows LF programs to
transparently scale to a given problem size without recompiling.

\begin{figure}[t]
  \centering
  \begin{subfigure}{.495\linewidth}
    \centering
    \includegraphics[scale=0.48]{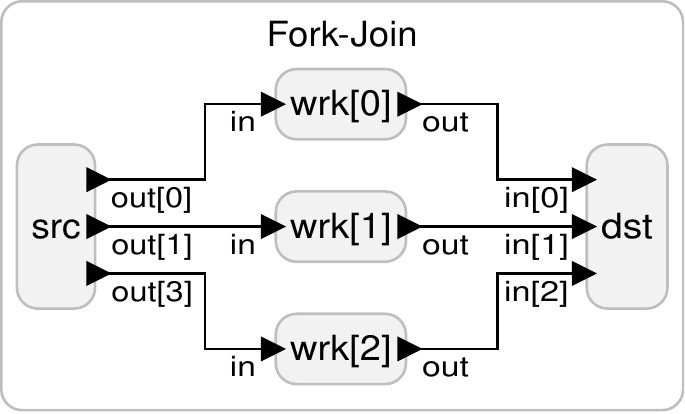}
    \caption{Multiport connections}
    \label{fig:rop:fork-join}
  \end{subfigure}\hfill
  \begin{subfigure}{.485\linewidth}
    \centering
    \includegraphics[scale=0.48]{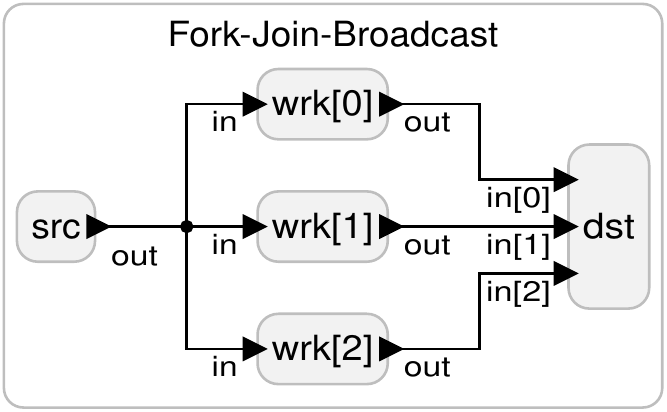}
    \caption{Broadcast connection}
    \label{fig:rop:fork-join-broadcast}
  \end{subfigure}
  \begin{subfigure}{\linewidth}
    \centering
    \includegraphics[scale=0.48]{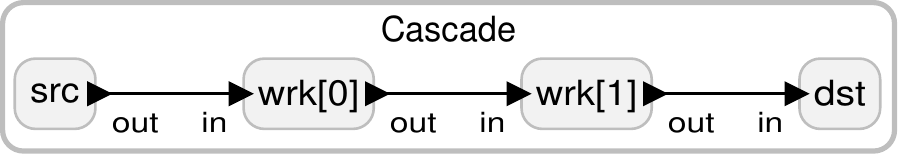}
    \caption{A cascade composition}
    \label{fig:rop:pipeline}
  \end{subfigure}
  \begin{subfigure}{\linewidth}
    \centering
    \includegraphics[scale=0.48]{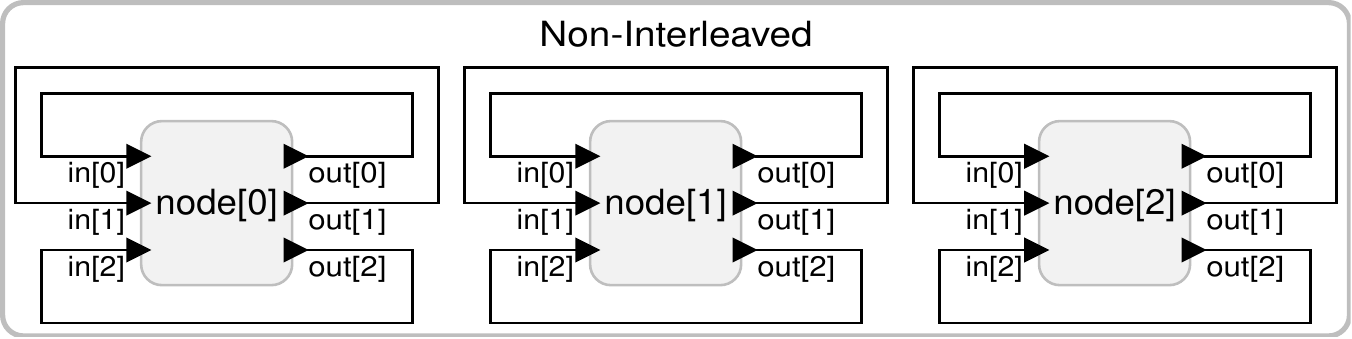}
    \caption{Multiports within a bank: direct connections}
    \label{fig:rop:non-interleaved}
  \end{subfigure}
  \begin{subfigure}{\linewidth}
    \centering
    \includegraphics[scale=0.48]{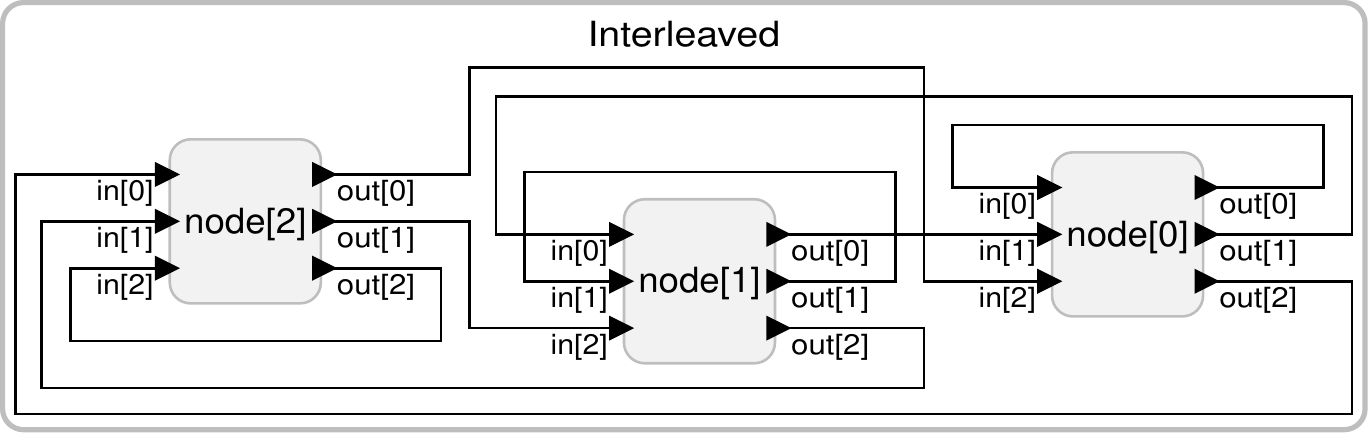}
    \caption{Multiports within a bank: interleaved connections}
    \label{fig:rop:interleaved}
  \end{subfigure}\\
  \vspace{-4pt}
  \caption{Various connection patterns in \lf}
  \vspace{-11pt}
\end{figure}

Consider a simple fork-join program in \lfshort:\footnote{
   Implementation details are omitted for
   brevity. Please refer to~\cite{LohstrohEtAl:21:Towards} for an introduction to
   LF syntax.
}\\
\begin{minipage}{0.52\linewidth}
  \begin{lstlisting}[language=LF,escapechar=|]
reactor Src(w: int(3)) {
  output[w] out: int
}
reactor Worker {
  input in: int
  output out: int
}
reactor Sink(w: int(3)) {
  input[w] in: int
}
  \end{lstlisting}
\end{minipage}%
\begin{minipage}{0.48\linewidth}
  \begin{lstlisting}[language=LF,firstnumber=11,escapechar=|]
main reactor(w: int(3)) {
  src = new Src(w = w) |\label{ln:out-width}|
  dst = new Sink(w = w)
  wrk = new[w] Worker() |\label{ln:worker-bank}|
  src.out -> wrk.in |\label{ln:fan-out}|
  wrk.out -> dst.in |\label{ln:fan-in}|
}
  \end{lstlisting}
\end{minipage}\\
The program defines a \code{Src}, a \code{Worker}, and a \code{Sink} reactor and
an unnamed \code{main} reactor that assembles the program.
\code{Worker} defines two individual ports of type \code{int} called \code{in} and
\code{out}.
\code{Src} and \code{Sink} use our syntax extension to each define a
\textbf{multiport} of width \code{w}, where \code{w} is a parameter and defaults
to 3.
The main reactor creates all reactor instances and connections.
Concretely, it creates two individual instances of \code{Src} and \code{Sink}
and uses our syntax extension to create a \textbf{bank} of worker reactors of
width \code{w} (line~\ref{ln:worker-bank}).
The two connection statements (line \ref{ln:fan-out}, \ref{ln:fan-in})
establish \code{w} connections each, one for each pair of multiport and bank
instance. The resulting connection pattern is illustrated in Figure~\ref{fig:rop:fork-join}.
Note that the precise number of workers is configurable via the \code{w}
parameter of the main reactor, which can be specified when executing the program without recompiling.
Hence the program can be configured to an arbitrary number of workers.

In this example, the source reactor produces three separate values
to be sent to the worker. Instead, if we want to broadcast a single value to all
workers, then we can use the broadcast syntax \code{(...)+}. Configuring
the source reactor to use a single output (by setting \code{w=1} in line~\ref{ln:out-width}) 
and changing line~\ref{ln:fan-out}
to \code{(src.out)+ -> wrk.in} creates the pattern in
Figure~\ref{fig:rop:fork-join-broadcast}.

Another common pattern that can be conveniently expressed using \lfshort syntax is a cascade
composition, illustrated by the following program:\\
\begin{minipage}{0.74\linewidth}
\begin{lstlisting}[language=LF,escapechar=|]
main reactor(n: int(2)) {
  src = new Src(w = 1)
  dst = new Sink(w = 1)
  wrk = new[n] Worker()
  src.out, wrk.out -> wrk.in, dst.in |\label{ln:after}|
}
\end{lstlisting}
\end{minipage}\\ 
The connection operator sequences all ports listed on the left- and right-hand side,
and connects the $n$th port on the left hand side to the $n$th port on the right-hand side.
By offsetting the left-hand side of the connection statement in
line~\ref{ln:after} with a single source port and appending the sink port to the
right hand side, we can effectively arrange the connections to form the cascade
shown in Figure~\ref{fig:rop:pipeline}.

The connection operator also connects multiports within banks. In
this case, the operator will implicitly unfold all port instances on both sides
of the connection to form a flat list of ports. The unfolding happens such that
we first list all ports of the first bank instance, then all ports of the second
instance, and so on. Consider the following program:\\
\begin{minipage}{0.48\linewidth}
  \begin{lstlisting}[language=LF]
reactor Node(w: int(3)) {
  input[w] in: int
  output[w] out: int
}
  \end{lstlisting}
\end{minipage}%
\begin{minipage}{0.48\linewidth}
  \begin{lstlisting}[language=LF,firstnumber=5,escapechar=|]
main reactor(w: int(3)) {
  node = new[w] Node(w=w)
  node.out -> node.in |\label{ln:interleaved}|
}
  \end{lstlisting}
\end{minipage}\\
This will create the pattern shown in Fig.~\ref{fig:rop:non-interleaved} which is not very useful. Using the
\code{interleaved} modifier on either side of the connection in
line~\ref{ln:interleaved}, we can modify the unfolding strategy to first list
all first port instances within all bank instances, then the second port
instances within all bank instances, and so on. This creates the fully connected
pattern shown in Figure~\ref{fig:rop:interleaved}.

All of the patterns discussed in this section are used extensively in our benchmark
implementations in Section~\ref{sec:evaluation}.

\subsection{Runtime Implementation}\label{sec:runtime}

The execution of each \lfshort program is governed by a runtime. Most importantly,
the runtime includes a scheduler which keeps track of all scheduled future
events, controls the advancement of logical time, and invokes any triggered
reactions in the order specified by the dependency graph while aiming to exploit
as much parallelism as possible.
Lohstroh et al. have already sketched a simple scheduling algorithm for reactor programs~\cite{Lohstroh:2019:CyPhy}.
In this section, we present a C++ implementation of this scheduling algorithm
that aims at exploiting parallelism while keeping synchronization overhead to a
minimum and avoiding contention on shared resources.

\begin{figure}[t]
  \scalebox{.7}{
  \begin{tikzpicture}[->,>=Latex,thick]
    \footnotesize
    \sffamily
    \node[align=center] (ie) {Initial-\\Events};
    \node[draw, right=2.5em of ie, minimum width=8em, fill=white] (eq) {Event Queue};
    \node[draw, below=1.2em of eq, minimum width=8em, fill=white] (reactq) {Reaction Queue};
    \node[draw, below=1.2em of reactq, minimum width=8em, fill=white] (readyq) {Ready Queue};
    \node[draw, right=8em of reactq, minimum width=7em, minimum height=7em, fill=white] (w) {Workers};
    \begin{scope}[on background layer]
      \node[draw, thick, minimum width=7em, minimum height=7em, fill=white] at ([yshift=9pt, xshift=9pt]w) {};
      \node[draw, thick, minimum width=7em, minimum height=7em, fill=white] at ([yshift=6pt, xshift=6pt]w) {};
      \node[draw, thick, minimum width=7em, minimum height=7em, fill=white] at ([yshift=3pt, xshift=3pt]w) {};
    \end{scope}
    \draw[->] (ie) to (eq);
    \draw[->] (eq) to (reactq);
    \draw[->] (reactq) to (readyq);
    \draw[->] (readyq) to (w.west |- readyq);
    \draw [->] (w) to node[above]{\texttt{set()}} (reactq);
    \draw [->] (w.west |- eq) to node[above]{\texttt{schedule()}} (eq);
  \end{tikzpicture}
  }
  \vspace{-8pt}
  \caption{Scheduling mechanism in the \lfshort runtime.}
  \label{fig:scheduler}
  \vspace{-10pt}
\end{figure}
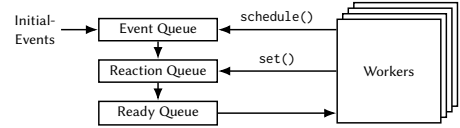

Figure~\ref{fig:scheduler} gives an overview of the scheduling mechanism used in our runtime.
The scheduler keeps track of future events in the \emph{event queue} and processes them
strictly in tag order.
When processing an event, the scheduler first determines all reactions that are
triggered by the event and stores them in the \emph{reaction queue}.
Any reactions in the reaction queue for which all dependencies are met (as indicated by the APG) are
forwarded to the ready queue and then picked up for execution by the worker threads.
If the executed reactions trigger any further reactions by setting ports, those reactions
are inserted in the reaction queue.
If a reaction schedules future events via an action, these new events are
inserted into the event queue.
Note that the scheduler always waits until all reactions at the current tag are
processed before advancing to the next tag and triggering new reactions.

The most important task of the scheduler is to decide when any given
reaction should be moved from the reaction queue to the ready queue. As the APG precisely defines the ordering constraints of reactions,
reaction scheduling is closely related to DAG-based scheduling
strategies~\cite{Kwok1999,Alam2018}.
However, the APG is not equivalent to a task graph
as it may contain reactions that do not need to be executed. Most
often only a fraction of the reactions is triggered at a particular tag.
Moreover, we do not know in advance precisely which reactions will be triggered
for a given tag, as reactions may or may not send
messages via their declared ports.
In consequence an optimal schedule cannot be computed in advance.

To decide whether a given triggered reaction is ready for execution,
we need to check if it has a dependency on any other reaction that is
triggered or currently executing.
To avoid traversing the APG at runtime, we utilize a simple heuristic.
Concretely, we assign a \emph{level} (top level as defined in~\cite{Kwok1999}) to each reaction.
Any reactions with the same level do not depend on each other and
hence can be executed in parallel.
Our scheduler then processes reactions going from one level to the next. Once all
reactions within a level are processed, all triggered reactions in the next
level are moved to the ready queue.
This approach avoids the need for analyzing the APG during execution, but also falls short on exploiting all
opportunities for parallel execution. For instance, this approach does not
execute reaction 2 of \texttt{ProxyDelay} in
parallel with reaction 2 of \texttt{Account}. Nonetheless, our evaluation shows
that this strategy is sufficient to efficiently exploit parallelism in
most cases. Given the extensive research on DAG-scheduling, we are confident
that we can apply more complex strategies in future work to also account for the
missed opportunities for exploiting parallelism.

Another limitation of our scheduling approach is that the scheduler only
executes reactions that are triggered at the same tag. In particular, this may
hinder exploiting pipeline parallelism in programs that do not use logical delays
to create separate pipeline stages. However, this limitation can be overcome by
using a federated execution strategy as discussed by \citet{Xronos}.

While the scheduling algorithm sketched in~\cite{Lohstroh:2019:CyPhy} and
refined in this section is relatively straight forward to implement, further
optimizations where needed to achieve competitive performance. In the following,
we detail the most important optimizations that we use in our C++ runtime.

\paragraph{Coordinating worker threads.}
In the above discussion we conceptually distinguished the scheduler from the
workers. In an actual implementation, however, using a central scheduler and
separate worker thread imposes a significant synchronization overhead. Instead,
in our implementation, any of the worker threads can become the scheduler and
move ready reactions to the ready queue or advance logical time to the next tag
if all reactions have been processed. Furthermore, we exploit the fact that at
any time we know the number of reactions to execute in parallel and use a
counting semaphore to control the number of active workers.

\paragraph{Lock-free data structures}
The three queues and other data structures that are required for bookkeeping
(e.g., a list of all set ports) are shared across workers.
Using mutexes for synchronization proved to be inefficient due to high
contention on the shared resources, especially when many parallel reactions set
ports or schedule actions.
Instead, we utilize lock-free data structures where possible.
For instance, the ready queue is implemented as a fixed size buffer paired with
an atomic counter. Since we know precisely how many reactions can at most run in
parallel (i.e. the maximum number of reactions in the APG that have the same
level), we can fix the size of the queue.
Every time new reactions are moved to the reaction queue, the atomic counter is
set to the number of reactions in the queue.
Each time a worker tries to execute a reaction it atomically decrements the counter. If the
result is negative, then the queue is empty. Otherwise the result provides the
index within the buffer to read from.

We further exploit knowledge about the execution of reactor programs. For instance,
the scheduler advances logical time only once all reactions have been processed.
This operation is safe without additional synchronization, as all of the workers are waiting for new reactions.

\paragraph{Sparse multiports.}
In programs where reactions in multiple reactors may trigger the same reaction
(such as an account with an arbitrary number of users), the triggered reaction
often needs to know which port(s) actually are set (contain data). If there are many upstream
reactors and communication is sparse, simply checking all ports for presence can
be inefficient. Instead, we expose an API for obtaining an iterator to only set
ports.
Note that this problem does not arise
in actors, as no ports exist and messages are processed one by one, only
considering those that are actually sent.

\section{Performance evaluation}
\label{sec:evaluation}

The actor model is widely accepted for programming large concurrent applications,
and implementations such as the C++ Actor Framework (CAF)~\cite{chs-rapc-16} and
Akka~\cite{AkkaAction2016} are known to be fast and efficient in utilizing a
larger number of threads. Compared to actors, \lfshort imposes various restrictions
that amount to a model of computation in which fewer behaviors are allowed.
In this section, we show
that these restrictions do not necessarily introduce overhead or higher
execution times. In fact, \lfshort is considerably faster for many benchmarks.

\subsection{Methodology}
Our evaluation is based on the Savina benchmark suite~\cite{savina} for actor
languages and frameworks. While this suite has several issues, as Blessing et
al. discuss in~\cite{blessing-19-run-actor-run}, Savina covers a wide range of
patterns and, to the best of our knowledge, is the most comprehensive benchmark suite
for actor frameworks that has been published. The Savina suite includes
Akka implementations of all benchmarks. CAF implementations of most Savina
benchmarks are also available.\footnote{\url{https://github.com/woelke/savina}}

We ported 22 of the 30 Savina benchmarks to the C++
target of \lfshort.\footnote{Source code available at \url{https://github.com/lf-lang/benchmarks-lingua-franca}}
Due to the fundamental differences between the actor and reactor model, the
process of porting benchmarks is not always straightforward. We aimed at closely
resembling the original workloads and considered the intention behind the
individual benchmarks.
We did not implement the benchmarks Fork Join (actor creation), Fibonacci,
Quicksort, Bitonic Sort, Sieve of Eratosthenes, Unbalanced Cobwebbed Tree,
Online Facility Location, and Successive Over-Relaxation as they require the
capability to dynamically create actors.
In the reactor model, this can be achieved with
mutations that may modify the reactor topology~\cite{Lohstroh:2019:CyPhy, Lohstroh:EECS-2020-235}.
However, mutations are not yet fully implemented in \lfshort, and a discussion of
language-level constructs for supporting mutations is beyond the scope of this
paper.

We further omit the A*-Search and Logistic Map Series benchmarks from our presentation.
A*-Search suffers from a severe race condition that results in wildly varying
execution times~\cite{blessing-19-run-actor-run}.
Logistic Map Series is omitted as it violates actor semantics and requires explicit synchronization~\cite{blessing-19-run-actor-run}. For this reason,
the CAF implementation needs to use a blocking call, which makes it slower than
the other implementations by at least two orders of magnitude.
Since this is not a problem of CAF, but rather a problem in the benchmark
design, we omit Logistic Map Series to avoid skewing the analysis.

\begin{figure*}[tb]
  \includegraphics[width=\textwidth]{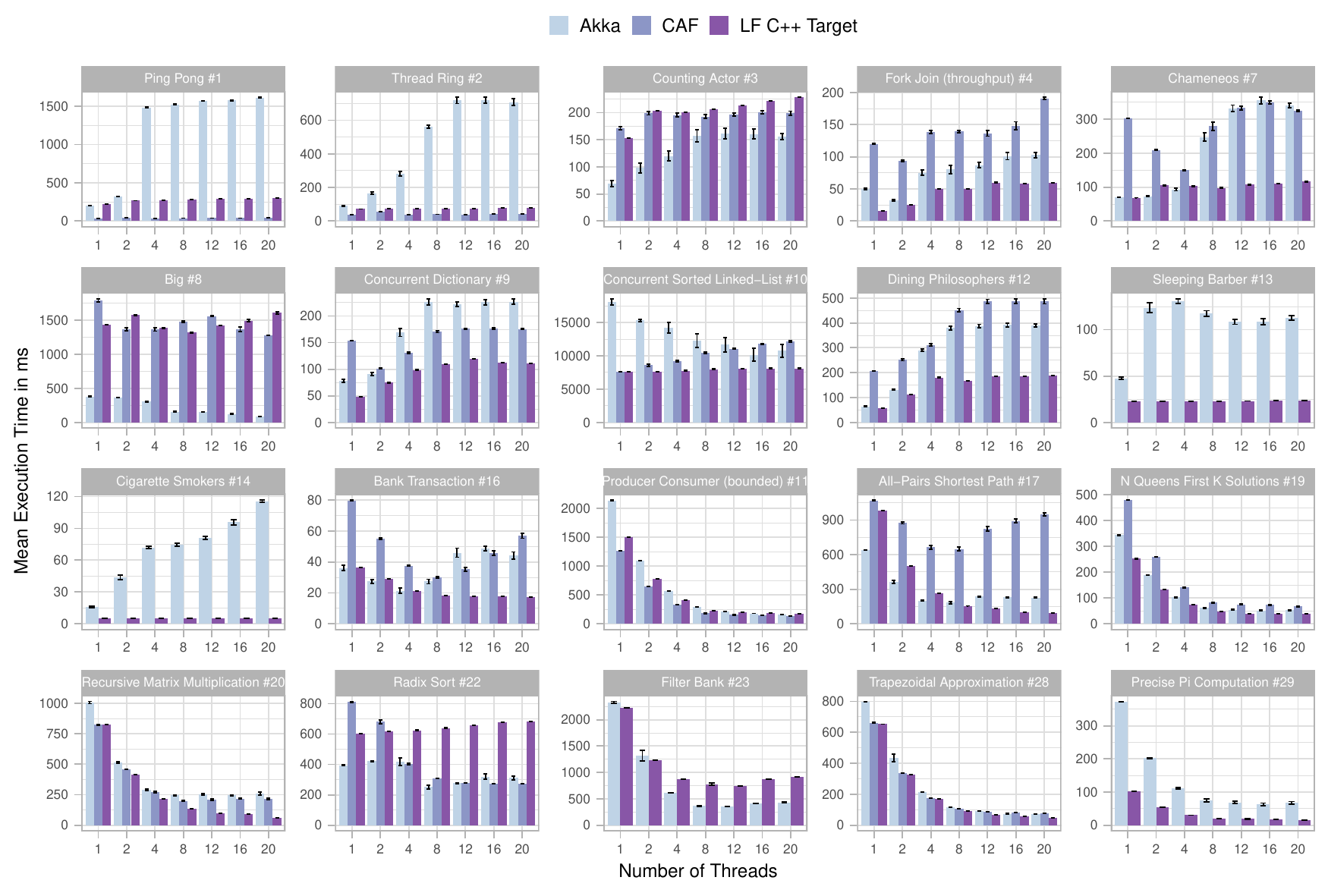}%
  \vspace{-12pt}%
  \caption{Mean execution times and 99\% confidence intervals for various Savina benchmarks implemented in \lfshort, CAF, and Akka, measured for a varying number of worker threads. The numbers prefixed with \# are benchmark IDs as listed in~\cite{savina}.}
  \label{fig:evaluation:results}
  \vspace{-6pt}
\end{figure*}

Figure~\ref{fig:evaluation:results} reports measured results for all supported benchmarks obtained with Akka, CAF,
and the C++ target of \lfshort.
The plots show the mean execution
times (including 99\% confidence intervals) for a varying number of worker threads for each of the benchmarks.
Not all benchmarks are implemented in CAF and hence it is missing in some plots.

All measurements were performed on a workstation with an Intel Core i9-10900K
processor with 32 GiB DDR4-2933 RAM running Ubuntu 22.04 and using CAF version
17.6 and Akka version 2.6.17. Following the methodology of Savina, measurements
exclude initialization and cleanup. Each measurement comprises 32 iterations.
The first two iterations are excluded from our analysis and are used to warm up
caches and the JVM (in the case of Akka).

\subsection{Discussion}

The first six plots in Figure~\ref{fig:evaluation:results} belong to the group of
micro benchmarks in the Savina suite. These are designed to expose overhead
in the protocol used for exchanging messages and for scheduling. Overall,
our C++ runtime shows comparable performance to Akka and CAF.
In Ping Pong and Thread Ring, our implementation is considerably faster than Akka but is still outperformed by CAF.
For Counting Actor and Big, Akka performs better and the LF performance is slightly behind CAF.
In Fork Join and Chameneos, the \lfshort implementation outperforms both Akka and CAF, especially when using a larger number of worker threads.

The next six plots (Concurrent Dictionary to Bank Transaction) belong
to the group of concurrency benchmarks. \lfshort
significantly outperforms CAF and Akka in all the concurrency benchmarks (especially for a high number of worker threads). This highlights how concurrent
behavior is expressed naturally in \lfshort and can be executed efficiently.
Actor implementations of those benchmarks, on the other hand, have to synchronize explicitly
and resort to potentially expensive protocols (e.g., by sending acknowledge messages), or implement
some other (blocking) synchronization strategy that violates actor
semantics~\cite{blessing-19-run-actor-run}.

The remaining plots belong to the group of parallelism benchmarks in the Savina
suite\footnote{Producer Consumer is actually listed as a concurrency benchmark,
but we find it fits better to the group of parallelism benchmarks.}.
Radix Sort and Filter Bank suffer somewhat from an inefficiency in our scheduler,
as discussed in Section~\ref{sec:runtime}.
In these particular
benchmarks, our simple algorithm leads to a non-optimal execution as some
reactions are executed later than they could. We will revise this algorithm in
future work.
However, the remaining parallelism benchmarks highlight that \lfshort can efficiently implement
parallel algorithms. Our \lfshort implementations are on par with Akka and CAF
and scale well with more threads. 

On average, \lfshort outperforms both Akka and CAF. For 20 threads, the C++
runtime achieves a speedup of \(1.85x\) over Akka and a \(1.42x\) speedup over
CAF. These speedups were calculated using the geometric mean over the speedups
of individual benchmarks. We conclude that \lfshort can compete with and even
outperform modern and highly optimized actor frameworks such as Akka and CAF. Particularly with workloads that require synchronization, \lfshort significantly
outperforms actor implementations. \lfshort is as efficient as the actor
frameworks in exploiting parallelism and scales well to a larger thread count.
In summary, the deterministic concurrency provided by \lfshort does not hinder performance but
enables more efficient implementations.
This is possible in part because the scheduler has insights into the program structure, and
explicit synchronization can be avoided in \lfshort, as opposed to many of Savina actor benchmarks.

The performance comparison between C++ and Scala (Akka)
needs to be taken with care, as other factors such as different library
implementations and the behavior of the JVM may influence performance. For
instance, the large discrepancy between Akka and our implementation in the Pi
Precision benchmark is explained by a less efficient representation of large
numbers in Scala/Java. However, the other benchmarks of the Savina suite do not depend on external libraries
and are designed to be more portable between languages.
Also note that over all benchmarks CAF only
achieves an average speedup of \(1.09x\) over Akka for 20 threads and is
outperformed in 9 out of 16 benchmarks.
For single threaded execution, Akka outperforms CAF in 10 benchmarks and
achieves an average speedup of 1.33x.
This indicates that the implemented Scala workloads
are comparable to the C++ implementations. Even considering a potential skew due to
the JVM, our results clearly show that \lfshort can compete with state-of-the-art actor frameworks.

\begin{figure}[t]
  \centering
  \includegraphics{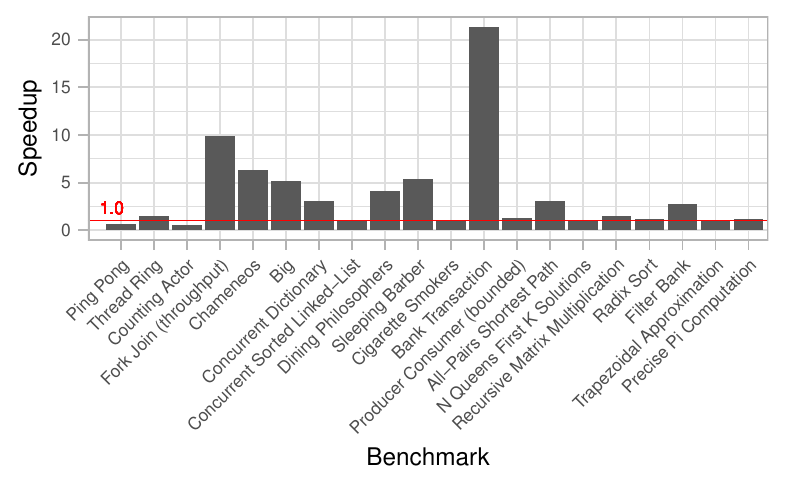}%
  \vspace{-12pt}%
  \caption{Speedup achieved by our optimized C++ runtime for 20 worker threads compared to an
    unoptimized version.}
  \label{fig:optimization}
  \vspace{-10pt}
\end{figure}

To better understand the impact of the optimizations discussed in
Section~\ref{sec:runtime}, Figure~\ref{fig:optimization} also shows the speedup
of our runtime for 20 worker threads compared to a less optimized runtime. This
baseline is an older version of our runtime that is optimized in the sense that
we used code profiling to identify obvious bottlenecks and eliminated them using
common code optimization techniques, but that does not include the optimizations
discussed in this paper. The average overall speedup (geometric mean) achieved
by our optimizations is \(2.18x\). In particular, Big and Bank Transaction
significantly benefit from our optimization for sparse communication patterns.
The concurrency benchmarks (e.g., Concurrent Dictionary and Dining
Philosophers), are mostly improved by reducing the contention on shared
resources. However, not all benchmarks benefit from our optimizations. The
reduced performance in Ping Pong and Counting Actor shows that optimizing for
efficient parallel execution also comes at a cost for simple sequential
programs.

\section{Related Work}
\label{sec:related}

\lfshort is closely related to the languages and frameworks evolved around
Hewitt's actor model~\cite{Agha:97:ActorComputation,hewitt2010actor}, including 
Akka~\cite{AkkaAction2016}, CAF~\cite{chs-rapc-16},
Ray~\cite{DBLP:journals/corr/abs-1712-05889}, Erlang~\cite{Armstrong:96:Erlang},
Rebeca~\cite{DBLP:journals/fuin/SirjaniMSB04}, P~\cite{DesaiEtAl:12:P} and Pony \cite{pony-actor-lang}.
Also reactive programming techniques~\cite{Bainomugisha:2013:SRP:2501654.2501666}, as used in 
frameworks like ReactiveX~\cite{Meijer:2010:REC:1900160.1900173} and Reactors.IO~\cite{Prokopec2018} but also
in language-level constructs like event loops~\cite{tilkov2010node}, are closely related to \lfshort.
While actors and reactive programming provide good resiliency and scalability,
this comes at the cost of nondeterminism, which makes programs notoriously hard
to test and debug~\cite{banken2018debugging,TorresLopez2019}.
Even more problems arise if languages, frameworks, and libraries do not enforce
the underlying model and invite the programmer to break its
semantics~\cite{tasharofi2013scala}. Pony addresses the later problem by
leveraging a strong type system similar to Rust to prevent data races at compile
time. Rebeca provides a formalism and model checking techniques for analyzing and verifying
actor networks. While this can improve confidence in a correct implementation,
the programmer is still responsible for finding this correct implementation.
P goes a step further in that it also has an efficient runtime system and
a compiler that generates correct-by-construction code with reasonable performance.

Blessing et al. propose a strategy that maps actor communication to a tree
topology in order to guarantee a causal ordering of
messages~\cite{Blessing2017}. In a similar approach, Sang et al. utilize
a DAG topology to achieve serializability in the processing of events.
Orleans~\cite{Bykov2011} is also based on an actor-like model and provides
guarantees on atomicity on transactions. Finally, Reactors as defined by~\citet{Field2009} is a
model that is closely related to actors but that supports both synchronous and
asynchronous communication and also provides atomicity guarantees. All these
strategies are most useful in distributed scenarios, in particular in presence
of network failures. In this paper, however, we focus on the execution of a
single host. Moreover, the determinism guarantees that \lfshort makes are
stronger. Nonetheless, such techniques are highly relevant to \lfshort and could be
deployed for ensuring fault-tolerant execution in distributed \lfshort programs. We
believe, however, that \lfshort provides a more general solution, as the
programmer can explicitly trade consistency for \mbox{availability} in distributed
contexts~\cite{CALTheorem}, and hence the solution can be adjusted to the
concrete application requirements.

Dataflow models~\cite{Dennis:74:Dataflow,Bilsen:94:CSDF,Lee:87:SDF} and process
networks~\cite{Kahn:74:PN,Lee:95:ProcessNetworks} provide deterministic
concurrency by creating statically connected networks of actors with
deterministic semantics. These
models enable improved static analysis and
optimization~\cite{Geilen:05:ModelChecking}, but they limit the
application's flexibility and capability to react to external events.
Ohua~\cite{ohua} is another language
providing parallelism through dataflow and is similar to \lfshort in that it
integrates with existing high-level languages. However, it falls short on
exposing coordination facilities for individual nodes and does not provide a timed semantics.

Deterministic concurrency is also found in synchronous languages such as
Esterel~\cite{Berry:92:Esterel}, Lustre~\cite{halbwachs91:lustre},
and SIGNAL~\cite{Benveniste:90:Signal} as well as in 
Functional Reactive Programming (FRP) languages, like Fran~\cite{elliott1997fran},
FrTime~\cite{cooper2006embedding}, and
Elm~\cite{czaplicki2013elm}.
However, these languages are challenging to use for general purpose programming as they require pure functions and
there is a lack of widely-applicable libraries.
Only recently, side effects have been considered in a formal semantics for Esterel~\cite{florence2019calculus} and
distributed dataflow~\cite{mogk2019fault}.
In \lf, arbitrary code can be embedded in reactions and we can benefit from the
available libraries for popular general purpose languages.

Another interesting approach is taken by deterministic multithreading libraries
such as DThreads~\cite{Liu2011} or Consequence~\cite{Merrifield2015}, which enforce   a
total order for concurrent store operations. Recent work has made considerable progress in 
avoiding the bottlenecks of conventional DTM techniques~\cite{Merrifield2019}. However,
we argue that threads are not a convenient concurrency model for the reasons outlined in~\cite{Lee:06:Threads}.
Moreover, threads do not allow for transparent distributed execution as is possible with (re)actors.

\section{Conclusion}
\label{sec:conclusion}

Unlike actors and related models for asynchronous concurrency, \lfshort enforces
determinism by default, and features asynchronous behavior only when introduced
deliberately. Our evaluation, based on \lfshort's C++ target, shows that this
deterministic model does not impede performance. On the contrary, we achieve an
average speedup of \(1.85x\) over Akka and \(1.42x\) over CAF. With \lfshort, we
manage to combine reproducible (and testable) behavior with good performance.
Yet, our relatively simple scheduling strategy %
likely
still leaves room for significant improvement. We leave it as
future work to explore more advanced scheduling algorithms capable of exploiting
more parallelism at runtime. We also aim to furnish full runtime support for
mutations and implement the remaining Savina benchmarks that require them. 
Finally, we note that our implementation of the Savina benchmark suite is not
only useful for comparing \lfshort to actor-oriented frameworks; it also demonstrates
that \lfshort, which is still in its infancy, is already suitable for
solving practical problems.

\begin{acks}
  The work in this paper was supported in part by the German Federal Ministry of
  Education and Research (BMBF) as part of the Software Campus (01IS12051) and
  the program ``Souverän. Digital. Vernetzt.'', joint project 6G-life (16KISK001K).
  This work was also supported in part
  by the National Science Foundation (NSF), award \#CNS-1836601 (Reconciling Safety with the Internet),
  the iCyPhy Research Center (Industrial Cyber-Physical Systems), supported by Denso, Ford, Siemens, and Toyota,
  and the National Research Foundation (NRF) of Korea (No. NRF-2022R1F1A1065201).
\end{acks}

\bibliography{../Refs,../msbib,support/Refs.bib}

\end{document}